\documentclass[aps,pra,superscriptaddress,nofootinbib,showpacs,doublespacing,preprint]{revtex4-1}

\usepackage{amsfonts}
\usepackage{amsmath}
\usepackage{graphicx}
\usepackage[sort&compress]{natbib}
\usepackage{hyperref}
\usepackage{braket}
\usepackage{overpic}
\usepackage{tikz}
\usetikzlibrary{matrix}
\usepackage{graphicx}

\begin{document}
\title{Exciton absorption in narrow armchair graphene nanoribbons}
\author{ B.~S.~Monozon}
\affiliation{Physics Department, Marine Technical University, 3 Lotsmanskaya Str.,\\
190008 St.Petersburg, Russia}
\author{P.~Schmelcher}
\affiliation{Zentrum f\"ur Optische Quantentechnologien, Universit\"{a}t Hamburg, \\ Luruper Chaussee 149, 22761 Hamburg, Germany}
\affiliation{The Hamburg Centre for Ultrafast Imaging, Universit\"{a}t Hamburg,~\\ Luruper Chaussee 149, 22761 Hamburg, Germany}
\date{\today}

\begin{abstract}
We develop an analytical approach to the exciton optical absorption
for narrow gap armchair graphene nanoribbons (AGNR). We focus on the regime
of dominant size quantization in combination with the attractive
electron-hole interaction. An adiabatic separation of slow and fast motions
leads via the two-body Dirac equation to the isolated and coupled
subband approximations. Discrete and continuous exciton states
are in general coupled and form quasi Rydberg series of purely discrete
and resonance type character. Corresponding oscillator strengths and
widths are derived. We show that the exciton peaks are blue-shifted,
become broader and increase in magnitude upon narrowing the ribbon.
At the edge of a subband the singularity related to the 1D density of states
is transformed into finite absorption via the presence of the exciton.
Our analytical results are in good agreement with those
obtained by other methods including numerical approaches. Estimates
of the expected experimental values are provided for realistic AGNRs.
\end{abstract}

\maketitle

\section{Introduction}\label{S:intro}

Spatially confined elongated strips of graphene monolayer
termed graphene nanoribbons (GNR) have attracted in recent years
substantial interest both
theoretically and experimentally
(see \cite{ Wakab, Alf, Soavi} and references therein).
GNR are of fundamental importance for nanoscience and nanotechnology applications.
In general, surpass the gapless 2D graphene monolayers with fixed
electronic, optical, and transport properties, and demonstrate
flexible features because of an open tunable electronic band gap
governed by the ribbon width.
In contrast to zigzag GNR, the armchair GNR (AGNR), which possess extrema
of the energy bands at the common centre of the Brillouin zone are more
amenable to a theoretical description. Below we will focus on the
semiconductor-like quasi-1D AGNR having open band gaps
$\Delta_{N_{eh}} = \varepsilon_{N_{e}} - \varepsilon_{N_{h}}$
determined by the distances between the
electron $(e)$ and hole $(h)$ size-quantized energy levels
$\varepsilon_{N_{e,h}} \sim d^{-1}$ induced
by a finite ribbon width $d$.

Optical absorption caused by the transitions between the electron and hole
subbands associated with the energy levels $\varepsilon_{N_{e,h}}$
represents an effective
tool to explore the electronic structure of the AGNR electronic structure.
The key point is the inverse
square-root divergence of the density of 1D states of free carriers at the
band gaps, manifesting itself in the inter-subband optical effects. However,
in the experimental spectra of real AGNR these singularities are replaced
by a more complicated pattern. Excitons
formed by the attractively interesting electron and
hole drastically change the optical absorption
properties in the vicinity of the edges
determined by the energy gaps.
It was shown for exciton absorption
in a bulk semiconductor subject to a strong magnetic field \cite{Hashow}
and in a narrow semiconductor quantum wire \cite{Monschm09} that
Rydberg series of exciton peaks arise below each edge and tend to a finite
absorption at edges thereby shadowing the square-root singularity and modifying
the fundamental absorption above the edges. In addition, quasi-1D semiconductor
structures are preferable for exciton studies. In units of the exciton
Rydberg constant $Ry^{\mbox{\tiny (x)}}$ the exciton binding energy $E_{b}$
in a 3D bulk material is $E_{b}^{(3\mbox{\tiny D})}=Ry^{\mbox{\tiny (x)}}$, in a 2D
quantum well $E_{b}^{(2\mbox{\tiny D})}=4Ry^{\mbox{\tiny (x)}}$, while in a 1D
quantum wire it is suppressed both
by a magnetic field or by the boundaries of the wire
$E_{b}^{(1\mbox{\tiny D})}\sim Ry^{\mbox{\tiny (x)}}\ln^2 (\frac{R}{a_{\mbox{\tiny x}}})$,~
($R<<a_{\mbox{\tiny x}}$), where $R$
is the wire radius or
the magnetic length. In this sense the AGNR surpass the semiconductor
structures for which the Rydberg constant is a fixed parameter,
while for the AGNR $Ry^{\mbox{\tiny (x)}}\sim \Delta_{N_{eh}} \sim d^{-1}$
\cite{Nov,Monschm12}.

To date, the discrete part of the exciton absorption spectrum
of GNR has been calculated numerically \cite{Yang, Prez, Zhu, Denk}
using approaches based
on density functional theory, the local density approximation, and the
Bethe-Salpeter equation.
Jia \emph{et al.} \cite{Jia} and Lu \emph{et al.} \cite{Lu} used
the tight-binding approximation, while Alfonsi and Meneghetti \cite{Alf}
employed the Hubbard Hamiltonian in their \emph{ab initio} calculations of the
positions and intensities of the exciton peaks. Only a few of the
recent approaches
\cite{Rat, Mohamm} relied on analytical methods based on the
nonrelativistic Wannier
1D model. Thus, by now the absolute majority of the
works focusing on the
problem of exciton absorption
in narrow AGNR are numerical
calculations aiming at the exciton binding energy or
and discrete exciton peaks.
A consistent analytical theory, that considers
the 2D two-body exciton Hamiltonian and gives rise
to quasi-1D bound and unbound excitons, inducing
discrete and continuous
optical absorption,
respectively, is virtually not addressed in the literature.
The inter-subband
interaction of the exciton states did not yet attract attention.
In addition,
the complete form of the exciton absorption coefficient has not been
derived explicitly. Undoubtedly, numerical calculations are required
for an adequate description of concrete experiments.
Nevertheless, analytical methods are indispensable to make the basic
physics of AGNR transparent
and then to promote the application of these materials in nano- and
opto-electronics using the dependence of the properties of the AGNR on
the ribbon width.

In order to fill the mentioned gaps we develop an analytical approach,
which yields an
explicit form of the exciton absorption coefficient
for AGNR.
The electron-hole Coulomb attraction is taken
to be much weaker than the effect of the ribbon confinement, which in turn
means a narrow ribbon as compared to the exciton Bohr radius. The two-body
Dirac equation describing the 2D massless electron-hole pair is solved
in the adiabatic approximation. This approximation implies that the transverse
motion of the particles governed by the ribbon confinement is much
faster than its longitudinal motion controlled by the 1D exciton field, which is
calculated by averaging the 2D
Coulomb exciton potential over the electron and hole transverse states.
In the single-subband approximation
of isolated $N$-subbands the exciton energy spectrum is
a sequence of series of the quasi-Coulomb strictly discrete $Nn$-
levels positioned below the $N$th size-quantized level and
continuous subbands originating from each $N$-level.
A coupling between the discrete and continuous exciton states,
specified by the $N=0,1$ levels is taken into account
in the double-subband
approximation. Inter-subband interaction converts the strictly discrete
$Nn$-states into the quasi-discrete ones (Fano resonances), having nonzero
energy widths, which manifest itself in the Lorentzian form
of the exciton absorption peaks.
Clearly, energetically lowest exciton series,
corresponding to the ground size-quantized level $N=0$
does not interact with the series lying above and remains discrete.

Our mathematical approach is based on matching the Coulomb wave
functions with the functions obtained upon solution of the Dirac
equation in the intermediate region by the iteration method.
This procedure has been originally developed
by Hasegawa and Howard \cite{Hashow} for 3D excitons subject to a strong
magnetic field and then successfully extended to problems related to
semiconductor (see \cite{Monschm09} and references therein) and graphene
\cite{Monschm12,Monschm14} nanostructures. The dependence of the
exciton absorption coefficient on the ribbon width is studied analytically. Our
results are in line with the conclusions based on numerical approaches
and the corresponding experimental data. The aim of this work is to make AGNR
attractive as optoelectronic devices due to
the strong dependence of the exciton spectrum on the ribbon width.

This work is organized as follows. In Section 2 the general
analytical approach is described. The energy levels of the
discrete states and the exciton wave functions
of the discrete and continuous energy spectrum are calculated
in Section 3 in the single-subband approximation.
The discrete exciton peaks and the continuous absorption
are considered in Section 4.
Section 5 provides the double-subband approximation
to study the optical series of the Fano $n$-resonances,
relevant to the first excited subband $N=1.$
In Section 6 we discuss the
obtained results, compare them with the available data, and estimate
the expected experimental values. Section 7 contains our conclusions.

\section{General approach}\label{S:gen}

Below we consider the exciton absorption in an AGNR with width $d$
and length $L$ placed on the $x-y$ plane and bounded by straight lines $x=\pm d/2.$
The polarization of the light wave is assumed to be parallel to the $y$-axis.
Optical absorption in GNR associated with electron
interband transitions has been studied numerically by Hsu and Reichl
\cite{Hsu} as well as Gundra and Shukla \cite{Gund}, while a comprehensive
analytical approach was recently developed by Sasaki \emph{et al} \cite{Sas}.
In particular,
it was shown that the inter-subband $y$-polarized
transitions are allowed between the electron $N_e$ and the hole $N_h$ subbands
with the same indices $N_e = N_h = N$.
Elliot justified thoroughly in Ref. \cite{Ell} that the exciton absorption
in semiconductors, can be treated
as the electron-hole pair optical transition from the
ground state described by the wave function
$\Psi^{(0)}(\vec{\rho}_e ,\vec{\rho}_h)$ to the excited exciton state
$\Psi^{\mbox \tiny (x)}(\vec{\rho}_e ,\vec{\rho}_h)$, in which the electron
$(e)$ and hole $(h)$ with coordinates $\vec{\rho}_e$ and $\vec{\rho}_h$
are in the conduction and valence band, respectively.  On extending the results of
Elliot \cite{Ell} and
Sasaki \emph{et al.} \cite{Sas} to exciton transitions
in a semiconductor-like AGNR,
the exciton absorption coefficient $\alpha$ becomes

\begin{equation}\label{E:abs}
\alpha = \sum_N \alpha^{(N)};\quad
\alpha^{(N)} = \frac{1}{n_b \varepsilon_0 c}\sigma_{yy}^{(N)},
\end{equation}
where $\sigma_{yy}^{(N)}$ is the component of the dynamical conductivity

\begin{equation}\label{E:cond}
\sigma_{yy}^{(N)}=  \frac{\pi p^2 e^2}{\hbar S \Delta_{N}}
\sum_{n,s}\big |\sigma^{(N)}_{xn(s)}\big |^2 \delta \left(\hbar \omega - E_{Nn(s)}\right)
\delta_{\vec{q}_{ph} \vec{K}}
\end{equation}
determined by the matrix element

\begin{equation}\label{E:matr}
\sigma^{(N)}_{xn(s)} = \left <\vec{\Psi}^{(0)}(\vec{\rho}_e ,\vec{\rho}_h)
|\hat{\sigma}_{xh}\bigotimes \hat{I}_e + \hat{I}_h \bigotimes \hat{\sigma}_{xe}|
\vec{\Psi}^{(\mbox{ex})}_{Nn(s)}(\vec{\rho}_e ,\vec{\rho}_h)   \right >
\end{equation}
of the Pauli matrix $\hat{\sigma}_{x}$ calculated between the ground
$\vec{\Psi}^{(0)}$ and exciton $\vec{\Psi}^{\mbox{\tiny (x)}}_{Nn(s)}$ wave vectors
of the bound $(n)$ and continuous $(s)$ states of the exciton, formed
by an electron and hole from the corresponding energy subbands
with the common index $N$. As usual, the symbol $\bigotimes$ denotes the tensor
product of the Pauli $\hat{\sigma}_{x}$ and unit $\hat{I}$ matrices. In eq.
(\ref{E:abs}) $n_b$ is the refraction index of the ribbon substrate,
 $c$ is the speed of light, while in eq. (\ref{E:cond})
$p=\hbar v_F,~(v_F = 10^6\,\mbox{m/s})$ is the graphene energy parameter
, $S=Ld$ is the area of the ribbon, $\Delta_N = 2\varepsilon_N$ is the
effective energy gap between the electron and hole subbands,
branching from the size-quantized levels $\pm \varepsilon_N$
in conduction and valence bands,
respectively. The $\delta$-functions in eq. (\ref{E:cond})
reflect the conservation laws in the system
formed by the absorbed photon with the energy $\hbar \omega$
and momentum $\hbar\vec{q}_{\mbox{\tiny {ph}}}$ plus emersed exciton of the energy
$E_{Nn(s)}$ and the total momentum $\hbar\vec{K}$.

Following Elliott's approach \cite{Ell} the wave function $\vec{\Psi}^{(0)}$
related to the ground state of the electron-hole pair
in an AGNR can be chosen in the
form

\begin{equation}\label{E:grfun}
\vec{\Psi}^{(0)}(\vec{\rho}_e ,\vec{\rho}_h)=\delta (y)\delta (x_e -x_h)
\left[\vec{\Phi}_A^{(0)}(x_e)\bigotimes\vec{\Phi}_A^{(0)}(x_h) +
 \vec{\Phi}_B^{(0)}(x_e)\bigotimes\vec{\Phi}_B^{(0)}(x_h)\right],
\end{equation}
where $y=y_e -y_h$ is the relative $y$-coordinate and

$$
\vec{\Phi}_A^{(0)}(x) =\frac{1}{\sqrt{2}}
\begin{array}{c}
\begin{Bmatrix}
-1 \\
0 \\
1\\
0
\end{Bmatrix}
\end{array};
\qquad
\vec{\Phi}_B^{(0)}(x) =\frac{1}{\sqrt{2}}
\begin{array}{c}
\begin{Bmatrix}
0 \\
1 \\
0\\
-1
\end{Bmatrix}
\end{array}.
$$
The exciton wave function $\vec{\Psi}^{\mbox{\tiny (x)}}$  can be found
by solving the equation

\begin{equation}\label{E:basic}
\hat{{\rm H}}_{\mbox{\tiny x}}\vec{\Psi}^{\mbox{\tiny (x)}}(\vec{\rho_e }, \vec{\rho_h })=
E\vec{\Psi}^{\mbox{\tiny(x)}}(\vec{\rho_e }, \vec{\rho_h }).
\end{equation}
In this equation

\begin{equation}\label{E:exham}
\hat{{\rm H}}_{\mbox{\tiny x}} = \hat{{\rm H}}_h (\hat{\vec{k}}_h)
\bigotimes\hat{I}_e + \hat{I}_h \bigotimes \hat{{\rm H}}_e (\hat{\vec{k}}_e)
+\hat{I}_h\bigotimes \hat{I}_e V(\vec{\rho_e } - \vec{\rho_h })
\end{equation}
is the traditional exciton Hamiltonian \cite{Mohen} formed by the electron
and hole Hamiltonians $\hat{{\rm H}}_j (\hat{\vec{k}}_j),~j=e,h$ corresponding
to the nonequivalent Dirac points\\
$\vec{K}^{(+,-)}= \pm K, 0;~(K = 4\pi / 3a_0 ,~a_0 = 2.46\, \mbox{{\AA}}$
is the graphene lattice constant). The Hamiltonian
$\hat{{\rm H}}_j (\hat{\vec{k}}_j)$ is represented by
\cite{Monschm12, Brfer}

$$
\hat{{\rm H}}_j (\vec{k}_j)= p \left(\begin{array}{cc} -\vec{\sigma} \hat{\vec{k}}_j
& 0 \\0 & \vec{\sigma}^{*} \hat{\vec{k}}_j \\
\end{array} \right) ;\qquad    \hat{\vec{k}}_j=-i\vec{\nabla}_j;
$$
containing the Pauli matrices $\sigma_{x,y}$, unit matrices $\hat{I}_j,~j=e,h$
and the 2D Coulomb potential of the electron-hole attraction

\begin{equation}\label{E:coul}
V(\vec{\rho}_e,\vec{\rho}_h)=
-\frac{e^2}{4\pi\varepsilon_0 \epsilon_{\mbox{\tiny eff}}\sqrt{(x_e-x_h)^2 +(y_e-y_h)^2 }}.
\end{equation}
Here $\epsilon_{\mbox{\tiny eff}}=\frac{1}{2}(1+\epsilon + \pi q_0)$ is the effective
dielectric constant determined by the static dielectric constant $\epsilon$ of the
substrate and by the parameter $q_0 = e^2/4\pi \varepsilon_0 p \simeq 2.2 $ \cite{Nov, Hwang}.

Further we choose the exciton wave function $\vec{\Psi}^{\mbox{\tiny (x)}}$ in the form

\begin{equation}\label{E:exvec}
\vec{\Psi}^{\mbox{\tiny (x)}}(\vec{\rho_e }, \vec{\rho_h })=
\sum_N \vec{\Psi}_N^{\mbox{\tiny (x)}}(\vec{\rho_e }, \vec{\rho_h })
\end{equation}
where
$$
\vec{\Psi}_N^{\mbox{\tiny (x)}}(\vec{\rho_e }, \vec{\rho_h })=
\vec{\Psi}_N (\vec{\rho_h })\bigotimes\vec{\Psi}_N (\vec{\rho_e }),
$$
and where

\begin{equation}\label{E:partvec}
\vec{\Psi}_{N}(\vec{\rho}_j)=
\frac{1}{\sqrt{2}}\left[ u_{NA}(y_j)\vec{\Phi}_{NA}(x_j) +
u_{NB}(y_j)\vec{\Phi}_{NB}(x_j)\right];\quad j=e,h.
\end{equation}
are the single particle wave-functions both related to the $N$ subband.
The exciton states consisting
of the electron and the hole associated with the different
$N_e \neq N_h$ subbands are optically inactive and can be excluded from
the expansion (\ref{E:partvec}) (see \cite{Monschm05} and references therein).

In equation (\ref{E:partvec}) the sublattice wave functions
$\vec{\Phi}_{NA,B}$ are as follows

$$
\vec{\Phi}_{NA}(x_j) =
\begin{array}{c}
\begin{Bmatrix}
-\varphi_N(x_j) \\
0 \\
\varphi_N^{\ast}(x_j)\\
0
\end{Bmatrix}
\end{array};
\qquad
\vec{\Phi}_{NB}(x_j) =
\begin{array}{c}
\begin{Bmatrix}
0 \\
\varphi_N(x_j) \\
0\\
-\varphi_N^{\ast}(x_j)
\end{Bmatrix}
\end{array};\quad j=e,h
$$
where the functions $\varphi_N(x_j)$ are represented by

\begin{equation}\label{E:trfun}
\varphi_N(x_j) =\frac{1}{\sqrt{2d}}\exp \left \{ {\rm i}\left [x_j\frac{\pi}{d}(N-\tilde{\sigma})-\frac{\pi}{2}
\left(N+ \left[ \frac{Kd}{\pi} \right]    \right)  \right ]\right \}.
\end{equation}
These wave functions form the single particle orthonormal wave functions \cite{Monschm12}

\begin{equation}\label{E:trvec}
\vec{\Phi}_{N} (x_j)=\frac{1}{\sqrt{2}}\left [ \vec{\Phi}_{NA} (x_j) + \vec{\Phi}_{NB} (x_j)  \right ]~;
\end{equation}
satisfying the equations

$$
\langle\vec{\Phi}_{N'B,A}|\vec{\Phi}_{NA,B}\rangle =0~;\quad
\langle\vec{\Phi}_{N'A,B}|\vec{\Phi}_{NA,B}\rangle =
\langle\vec{\Phi}_{N'}|\vec{\Phi}_{N}\rangle =\delta_{N'N}.
$$
The introduced wave functions obey the equations

\begin{eqnarray}\label{E:tren}
\left.
\begin{array}{c}
\hat{H}_j (\hat{k}_{jx},0)\vec{\Phi}_{N} (x_j)=\varepsilon_N \vec{\Phi}_{N}(x_j);\\
\hat{H}_j (\hat{k}_{jx},0)\vec{\Phi}_{NA,B} (x_j)=\varepsilon_N \vec{\Phi}_{NB,A}(x_j);\\
\varepsilon_N=|N-\tilde{\sigma}|\frac{\pi p}{d};~N=0,\pm1,\pm2,\ldots~;
\quad\tilde{\sigma}=\frac{Kd}{\pi}-\left[ \frac{Kd}{\pi}\right]~;\,j=e,h~;
\end{array}
\right \}
\end{eqnarray}
Below to be specific we will consider AGNR of the family sigma =
1/3, providing
a semiconductor-like gap structure, and we leave aside the case sigma = 0
corresponding to the metall-like gapless ribbon.

The components $\varphi_N(x_j),\,(j=e,h)$ in (\ref{E:trfun})
and energies $\varepsilon_N$ in eq. (\ref{E:tren})
have been derived from the boundary conditions

\begin{equation}\label{E:bound}
{\rm e}^{{\rm i}Kx_j}\varphi_{N}(x_j)-{\rm e}^{-{\rm i}Kx_j}\varphi^{*}_{N}(x_j) = 0
\quad \mbox{at}\quad x_j = \pm \frac{d}{2};~j=e,h,
\end{equation}
which provide that the
electron $(e)$
and hole $(h)$ states $\vec{\Phi}_{NA,B}$ multiplied by the factor
$\exp\left[{\rm i}\vec{K}^{(+,-)}\vec{\rho}_j \right]$, and the
exciton state $\vec{\Psi}^{\mbox{\tiny (x)}}$ (\ref{E:exvec}) vanish at
both edges of the A and B sublattices
(see Ref. \cite{Brfer, Castnet} for details).

Thus the wave functions $\vec{\Phi}_{N} (x_j)$ (\ref{E:trvec})
constitute the basis set related to the transversely
confined $x$-motion of free carriers with the size-quantized energies $\varepsilon_N > 0$
in eq. (\ref{E:tren}), while the wave functions
$\vec{\Phi}_{N\alpha}(x_h)\otimes\vec{\Phi}_{N\beta}(x_e),\quad \alpha,\beta =A,B$ with

$$
\left< \vec{\Phi}_{N\alpha}(x_h)\otimes\vec{\Phi}_{N\beta}(x_e)
\big|\vec{\Phi}_{N'\gamma}(x_h)\otimes\vec{\Phi}_{N'\delta}(x_e)\right> =
 \delta_{NN'} \delta_{\alpha\gamma} \delta_{\beta\delta}
$$
form an orthonormalized basis set for the calculation
of the exciton wave function
$\vec{\Psi}^{\mbox{\tiny (x)}}$
(\ref{E:exvec}) with the expansion coefficients $u_{N\alpha}(y_h)u_{N\beta}(y_e).$

Eq. (\ref{E:basic}) for $\vec{\Psi}^{\mbox{\tiny (x)}}$
defined by eqs. (\ref{E:exvec})-(\ref{E:partvec}) -
(\ref{E:tren}) leads to the set
of equations

$$
u_{N\alpha}(y_h) u_{N\beta}(y_e)= \frac{{\rm e}^{{\rm i}QY}}{\sqrt{L}}\xi_{N\alpha,\beta}(y)~;\,
\xi_{NAA}=\xi_{N1}~;\,  \xi_{NAB,BA}= \frac{1}{\sqrt{2}}(\xi_{N2}\pm \xi_{N3})~;\,
\xi_{NBB}=\xi_{N4};
$$
written in terms of the centre of mass
$Y=\frac{1}{2}(y_e + y_h)$ and relative $y = y_e - y_h$ coordinates

\begin{eqnarray}\label{E:set}
\left.
\begin{array}{l}
U_{N}(y,E)\xi_{N1} +\Omega_N^{*}(Q)\frac{1}{\sqrt{2}}\xi_{N2}
+\sqrt{2}p\frac{\partial}{\partial y}\xi_{N3} +\sum_{N'\neq N}V_{NN'}(y)\xi_{N'1}=0~;\\
U_{N}(y,E)\xi_{N4} + \Omega_N(Q)\frac{1}{\sqrt{2}}\xi_{N2} + \sqrt{2}p\frac{\partial}{\partial y}\xi_{N3}+
  +\sum_{N'\neq N}V_{NN'}(y)\xi_{N'4}=0~;\\
U_{N}(y,E)\sqrt{2}\xi_{N2} + \Omega_N(Q)\xi_{N1}  + \Omega_N^{*}(Q)\xi_{N4}
+ \sum_{N'\neq N}V_{NN'}(y)\sqrt{2}\xi_{N'2}=0~;\\
U_{N}(y,E)\sqrt{2}\xi_{N3} -2p\frac{\partial}{\partial y}\xi_{N1}
-2p\frac{\partial}{\partial y}\xi_{N4} + \sum_{N'\neq N}V_{NN'}(y)\sqrt{2}\xi_{N'3}=0~;
\end{array}
\right \}
\end{eqnarray}
where $U_{N}(y,E) =V_{NN}(y) - E,~\Omega_N(Q) = \Delta_N + {\rm i}pQ $,

\begin{align}\label{E:1Dc}
V_{N'N}(y)=\frac{1}{d^2}\int_{-\frac{d}{2}}^{+\frac{d}{2}}dx_e
\int_{-\frac{d}{2}}^{+\frac{d}{2}}dx_h V(\vec{\rho})
\cos \left[(N-N')\pi \left(\frac{x_e}{d}- \frac{1}{2}\right)  \right]
\\
\nonumber
\times
\cos \left[(N-N')\pi \left(\frac{x_h}{d}- \frac{1}{2}\right)  \right]~;\,\vec{\rho} = \vec{\rho}_e - \vec{\rho}_h~,
\end{align}
with eq. (\ref{E:coul}) for the potential $V(\vec{\rho})$. At $|y|\gg d$

\begin{equation}\label{E:onoff}
V_{N'N}(y)=-\frac{\beta}{|y|}\left[ \delta_{N'N}+O \left(\frac{d^2}{y^2}  \right)
\delta_{|N'-N|(2s+1)} \right]~;\,s=0,1,2,\ldots ;
\end{equation}
In eq. (\ref{E:set}) $\hbar Q$
is the longitudinal component of the exciton total momentum and $\Delta_N = 2\varepsilon_N$
is the effective energy gap between the electron and hole energy subbands.

Below we calculate the exciton states in the adiabatic approximation successfully
employed in the literature for impurity electron states in AGNR
\cite{Monschm12} and in electrically biased ribbons \cite{Monschm14}.
It is implied that the longitudinal $y$-motion is much slower
than the transverse $x$-motion i.e. the effect of the ribbon confinement significantly
exceeds that of the exciton attraction. The adiabaticity parameter $q$
representing the strength of the Coulomb potential in eq. (\ref{E:coul}) scaled
with the graphene energy parameter $p$ yields under the condition of adiabaticity

\begin{equation}\label{E:adiab}
q << 1,\quad \mbox{where}\quad q=\frac{e^2}{4\pi\varepsilon_0 \epsilon_{\mbox{\tiny eff}}p}.
\end{equation}
It allows us to set

$$
V_{NN'}(y),V_{NN'}^{'}(y),V_{NN'}^{''}(y) \ll E~;\,(E \simeq \Delta_N),\quad
\xi_{N1}\simeq \xi_{N4}\simeq \frac{1}{\sqrt{2}}\xi_{N2},\, \xi_{N3}\ll \xi_{N1}
$$
in eqs. (\ref{E:set})
and consequently reduce these equations to those for the functions $\xi_{N1}$,
which describe the nonrelativistic 1D exciton with reduced mass
$\mu =\frac{\hbar^2 |N - \tilde{\sigma} |\pi}{2pd}$
being governed by the quasi-Coulomb potentials $V_{NN'}(y)$ (\ref{E:1Dc}).

\section{Single-subband approximation}\label{S:Single}

Here we employ the single-subband approximation ignoring
the coupling between the reduced electron-hole subbands with the different
$N$. It follows from eq. (\ref{E:onoff}),
that in the narrow ribbon of small width $d$ the diagonal potentials $V_{NN}$
dominate the off-diagonal terms in the
set of equations (\ref{E:set}) almost everywhere
but for a small region $|y| < d$. This allows us to take
$V_{N'N}=V_{N}\delta_{N'N}$ and then to decompose the set (\ref{E:set})
related to the nonrelativistic
exciton into independent equations

\begin{equation}\label{E:geneq}
\xi_{N1}^{''}(y)
+ \frac{\left[-2EV_{NN}(y) + E^2  - \Delta_N^2 -p^2 Q^2 \right]  }
{4p^2 \left[1 - \frac{p^2 Q^2}{E^2} \right]}    \xi_{N1}=0
\end{equation}
with the diagonal potentials

\begin{eqnarray}\label{E:diag}
V_{N}(y)=\frac{2\beta}{d}\left[\ln\frac{\frac{|y|}{d}}{1 + \sqrt{1+\frac{y^2}{d^2}}}+
\sqrt{1+\frac{y^2}{d^2}} - \frac{|y|}{d}\right]=
\left\{
\begin{array}{cl}
\frac{\beta}{d}\ln\frac{y^2}{d^2}~;\, &\frac{|y|}{d}\ll 1\\
-\frac{\beta}{|y|}~;\, &\frac{|y|}{d}\gg 1
\end{array}
\right.
\end{eqnarray}
calculated from eq. (\ref{E:1Dc}).

\subsection{Exciton states}

The method of solving eq. (\ref{E:geneq})
 has been developed originally by Hasegawa
and Howard \cite{Hashow} in connection with
the problem of an exciton in a bulk semiconductor in the
presence of strong magnetic fields. The key point of their method is
the matching of the Coulomb wave function
corresponding to the potential $V_N(y) \sim -|y|^{-1}$ and that derived by
the iteration procedure using the exact potential $V_N(y)$ (\ref{E:1Dc}).
Since by now this method was widely and successfully employed
for the study of impurities and excitons in low-dimensional semiconductor
structures (see \cite{Monschm09} and references therein for details)
as well as graphene \cite{Monschm12} only an outline
of the calculations will be given below. At this stage
we neglect the effect of the longitudinal total
momentum of the exciton $Q=q_{ph}$.
The correction to the exciton binding energy caused by
this momentum will
be taken into account later on.
\\
\\
\emph{Discrete states}
\\
\\
For $|y| >> d$ eq. (\ref{E:geneq}) with the potential
$V_N(y) \sim -|y|^{-1}$ (\ref{E:diag})
gives for the normalized wave function $\xi_{N1}(y)$

\begin{equation}\label{E:couldisc}
\xi_{N1}(y) = A_{Nn}W_{\kappa, \frac{1}{2}}(\tau)~;
\end{equation}
where

\begin{eqnarray}
\tau &=& \nu y~;\quad\nu^2 =\frac{\Delta_N^2 - E^2}{p^2}~;\quad
\kappa = \frac{Eq}{2p\nu} = n+\beta_{Nn}~;\quad n=0,1,2,\ldots~;
\nonumber\\
A_{Nn}^2 &=& \frac{q\Delta_N }{(2n)^3 n!^2 p}~;n=1,2,3,\ldots~;
A_{N0}^2 = \frac{q\Delta_N }{4\beta_{N0}p}~;n=0,
\end{eqnarray}
$W_{\kappa, \frac{1}{2}}(\tau)$ is the Whittaker function \cite{Abrsteg}.

For $d << |y| << \nu^{-1} $ we employ the trial function
$\xi_{N1}^{(0)}(0) = c_0$ and for the derivative $\xi_{N1}^{(0)'}(0) = 0$,
generating an even wave function $\xi_{N1}(y) = \xi_{N1}(-y)$, to obtain

\begin{equation}\label{E:itdisc}
\xi_{N1}(y) =c_0\left[1-\kappa\tau \left(  \ln\frac{2\tau}{\nu d} - \frac{1}{2} \right) \right].
\end{equation}
Equating function (\ref{E:itdisc}) to that derived from eq. (\ref{E:couldisc})
and using the standard expansion of the Whittaker function for $\tau << 1$ \cite{Abrsteg},
we arrive at the equation for the quantum defect $\beta_{Nn}=\kappa - n$

\begin{equation}\label{E:Y}
Y_N(\kappa)=0,
\end{equation}
where

\begin{equation}\label{E:def}
Y_N(\kappa) = \frac{1}{\kappa - n} -\frac{1}{2\kappa} +\ln\frac{q|N-\tilde{\sigma}|\pi}{2\kappa}
+\psi(1+\kappa) + 2C -\frac{1}{2}
\end{equation}
and for the coefficient

\begin{equation}\label{E:zeron}
c_0(n) = A_{Nn} \Gamma(-\kappa + 1)^{-1},
\end{equation}
where $\psi(x)$ is the psi function (the logarithmic derivative of the gamma function
$\Gamma (x)$) and $C=0.577$ is the Euler constant. The exciton energy levels

\begin{equation}\label{E:discen}
E_{Nn} = \Delta_N \left(  1-\frac{q^2}{8(n+\beta_{Nn})^2} \right)~;
N=0,\pm 1, \pm 2, \ldots;\quad n=0,1,2,\ldots
\end{equation}
are the quasi-Rydberg series adjacent to the reduced size-quantized energy $N$-level
from low energies. Clearly the equations (\ref{E:couldisc}) and (\ref{E:discen})
are valid under the condition
$\nu d << 1$ with $\nu = \frac{q \varepsilon_{N} }{p\kappa}$,
which is ensured by the smallness of the adiabatic parameter $q<<1$.
\\
\\
\emph{Continuous states}
\\
\\
At $|y| >> d$ the wave function $\xi_{N1}(y)$ normalized to
$\delta(E-E^{'})$ becomes

\begin{equation}\label{E:coulcont}
\xi_{N1}(y) = D_{N s}\left[{\rm e}^{{\rm i}\Theta}W_{{\rm i}\zeta, \frac{1}{2}}(t)
+ {\rm e}^{-{\rm i}\Theta}W_{-{\rm i}\zeta, \frac{1}{2}}(-t)\right]~;
\end{equation}
where

\begin{eqnarray}
t = {\rm i}s y~;\quad s^2 &=&\frac{E^2 - \Delta_N^2 }{p^2}~;\quad
\zeta = \frac{Eq}{2ps}~;
\nonumber\\
D_{N s}&=&\frac{1}{2}\left(\frac{\zeta}{\pi q p}  \right)^{\frac{1}{2}}
{\rm e}^{-\frac{\pi\zeta}{2}}~;
\end{eqnarray}
and $\Theta$ is the corresponding phase.

The iteration procedure performed with the trial function
$\xi_{N1}^{(0)}(0) = c_1$; $\xi_{N1}^{(0)'}(0) = 0$,
leads in the region $d << |y|<< s^{-1}$ to the even wave function

\begin{equation}\label{E:itcont}
\xi_{N1}(y) =c_1\left[1-\zeta |t|\left(  \ln\frac{2|t|}{s d} - \frac{1}{2} \right) \right].
\end{equation}

Matching the function (\ref{E:itcont}) with that derived from eq. (\ref{E:coulcont})
for $|t| <<1$ we obtain the equation for the phase $\Theta$

\begin{equation}\label{E:phase}
\lambda_N(\zeta)- \frac{\pi}{1-{\rm e}^{-2\pi\zeta }}\cot \left(\Theta +\sigma  \right)=0
\end{equation}
where

\begin{equation}\label{E:lambda}
\lambda_N(\zeta) =
\frac{1}{2}\left[\psi (1+{\rm i \zeta}) + \psi (1-{\rm i \zeta}) \right]
+\ln \frac{|N-\tilde{\sigma}|\pi q}{2\zeta} +2C - \frac{1}{2} = 0~;
\end{equation}
and $\sigma (\zeta) = \mbox{arg} \Gamma (-{\rm i}\zeta)$.
The coefficient $c_1 (s)$ becomes

\begin{equation}\label{E:zeros}
c_1 (s) = -2D_{N s}\zeta^{-1}\sin \left( \Theta +\sigma\right)
\left(\frac{\zeta \sinh \pi\zeta}{\pi}  \right)^{\frac{1}{2}}~.
\end{equation}
The equations (\ref{E:coulcont}) and (\ref{E:phase})
are valid for the energy region above the threshold $\Delta_N$,
for which $sd \ll 1$.

\section{Spectrum of the exciton absorption}\label{S:spectrum}

Using eq. (\ref{E:grfun}) for the ground state wave function
$\vec{\Psi}^{(0)}$ and eqs.
(\ref{E:exvec})-(\ref{E:tren}) with the functions
$\varphi_N(x_j)$ multiplied by
the factors $\exp\left( \pm {\rm i}Kx_j \right)$
(see eq. (\ref{E:bound})), we calculate the matrix element
(\ref{E:matr}) of the dipole exciton optical transition in the form
$\sigma_{xn(s)}^{(N)}= -\sqrt{L}\xi_{N1}(0)$. As expected
for the noninteracting electron-hole pair for which
$\xi_{N1}(y)=\frac{1}{\sqrt{L}}{\rm e}^{{\rm i}sy}$
the matrix element of the fundamental optical transition
$\big|\sigma_{xs}^{(N)} \big|= 1$ coincides with that calculated in Ref. \cite{Sas}.
The contribution $\alpha^{(N)}$ (see eqs. (\ref{E:abs}), (\ref{E:cond})) to the
coefficient of the exciton absorption $\alpha$ in the vicinity of the edge
$\Delta_N = 2\varepsilon_N$ reads

\begin{equation}\label{E:absgen}
\alpha^{(N)}(\omega) = \alpha_0\frac{4\pi p^2}{n_b \Delta_N d}
\sum_{n(s)}\big| \xi_{N1}(0) \big|^2
\delta(\hbar \omega - E_{n(s)}^{(N)})~;
\end{equation}
where $\alpha_0 = e^2/4\varepsilon_0 \hbar c\simeq 2.3\cdot10^{-2}$
is the absorption of the suspended graphene. The coefficients
$\xi_{N1}(0)= c_0 (n)$ (eq. (\ref{E:zeron})) and
$\xi_{N1}(0)= c_1 (s)$ (eq. (\ref{E:zeros})) are responsible
for the oscillator
strengths of the discrete spectral peaks and for the shape
of the continuous absorption, respectively.
\\
\vspace*{1.5cm}
\\
\emph{Discrete spectrum}
\\
\\
It follows from eq. (\ref{E:absgen}) that the discrete spectrum
of the exciton absorption is
a Rydberg series with peaks at the frequencies

\begin{equation}\label{E:discsp}
\hbar\omega_{n}^{(N)} = \Delta_N - \frac{Ry^{\mbox{\tiny (x)}}}{(n+\beta_{Nn})^2}~;\,
n = 0,1,2,\ldots ,
\end{equation}
where $Ry^{{\mbox{\tiny (x)}}} = \frac{\hbar^2}{2\mu_{\mbox{\tiny x}}a_{\mbox{\tiny x}}^2},~
 a_{\mbox{\tiny x}}=\frac{4 \pi \varepsilon_0 \epsilon_{\mbox{\tiny eff}}\hbar^2}{\mu_{\mbox{\tiny x}}e^2},~
\mu_{\mbox{\tiny x}}=\frac{\hbar^2\varepsilon_N}{2p^2}~
\mbox{are the exciton Rydberg constant}$,~
Bohr radius and reduced mass, respectively.
All these parameters are induced by the ribbon confinement.

The oscillator strengths $f_n^{(N)}$ of the exciton $n$-peaks have the form

\begin{eqnarray}\label{E:discmax}
\frac{f_n^{(N)}}{L}= \big |c_0 (n)|^2 =
\frac{q\Delta_N }{4p}
\left\{
\begin{array}{cl}
\beta_{Nn}^2(2n^3)^{-1}~;\, &n=1,2,3,\ldots~;\\
\beta_{N0}^{-1}  ~;\, &n=0~;
\end{array}
\right.
\end{eqnarray}
where the quantum defects $\delta_{Nn}$ are given in eq. (\ref{E:def}).
\\
\\
\emph{Continuous spectrum}
\\
\\
In view of eq. (\ref{E:zeros}) the continuous spectrum of the
exciton absorption (\ref{E:absgen}) is given by

\begin{equation}\label{E:contmax}
\alpha^{(N)}(\omega)= \alpha_0\frac{4\pi p^2}{n_b \Delta_N d}
\big |c_1(s)|^2,
\end{equation}
with

$$
\big |c_1(s)|^2 = \frac{1}{pq} \pi \zeta Z(\zeta)
\frac{1}{\pi^2 +\lambda_N^2 (\zeta)\left(1- {\rm e}^{-2\pi\zeta} \right)^2}~;
$$
with $\zeta^2 = \frac{Ry^{\mbox{\tiny (x)}}}{\hbar\omega - \Delta_N }$. The function
$\lambda_N(\zeta)$ is given by eq. (\ref{E:lambda}).

In the vicinity of the edge $(\hbar \omega  = \Delta_N; \zeta \rightarrow \infty)$
we obtain from eqs. (\ref{E:absgen}) and (\ref{E:contmax})

\begin{equation}\label{E:edge}
\alpha^{(N)}(\omega)= \alpha_0\frac{2\pi p}{n_b \Delta_N d q}
\frac{1}{(\pi^2 + \lambda_{N\infty}^2)}
\left[  1 - \frac{\lambda_{N\infty}}{6(\pi^2 + \lambda_{N\infty}^2)\zeta^2}  \right]~;
\end{equation}
where

$$
\lambda_{N \infty} = \ln q + \ln \frac{|N-\tilde{\sigma}|\pi}{2}+2C - \frac{1}{2}.
$$

At the edge $(\zeta\rightarrow\infty)$ and in the logarithmic approximation
$|\ln q| >> 1$ eq. (\ref{E:edge}) reduces to

\begin{equation}\label{E:edge1}
\alpha^{(N)}(\omega)= \alpha_0\frac{1}{n_b|N-\tilde{\sigma}| q \ln^2 q},
\end{equation}
which in turn, as expected, coincides with the expression derived from
eqs. (\ref{E:absgen})-(\ref{E:discmax}) by replacing
$\sum_n \quad \mbox{by}\quad \frac{dn}{dE_n^{(N)}}
\quad \mbox{with}\quad \frac{dn}{dE_n^{(N)}}=
\frac{n^3}{2Ry^{\mbox{\tiny (x)}}}.$

The effect of the longitudinal total exciton momentum $Q$ can be calculated from
eq. (\ref{E:geneq}) and provides for the total energy $E_{Nn}$ of the bound exciton

$$
E_{Nn} = \Delta_N \left[1+ \left(\frac{pQ}{\Delta_N}\right)^2\right]^{\frac{1}{2}} -
 \frac{Ry^{\mbox{\tiny (x)}}}{(n+\beta_{Nn})^2}\left[1+ \left(\frac{pQ}{\Delta_N}\right)^2\right]^{\frac{3}{2}}~;
~n=0,1,2,\ldots~;
$$
where $ Ry^{\mbox{\tiny (x)}}$ is determined via eq. (\ref{E:discsp}).
Thus the motion of the centre of mass increases both the
total and binding energies of the exciton.

\section{Double-subband approximation}\label{S:Double}

At the next stage we take into account the coupling between the exciton states
of the discrete energy spectrum adjacent to the first excited
size-quantized energy gap $\Delta_1$ to states of the
continuous spectrum originating from the ground energy gap $\Delta_0$.
We set $N,N' = 0,1$ in the system of nonrelativistic equations

\begin{equation}\label{E:set1}
\xi_{N1}^{''}(y)
+ \frac{1}{4p^2}\left\{\left[-2EV_{N}(y) + E^2  - \Delta_N^2  \right] \xi_{N1}-
2E\Sigma_{N' \neq N} V_{NN'}(y)\xi_{N'1}\right\}  =0
\end{equation}
resulting from the general set (\ref{E:set}) for $q\ll 1$ and $Q=0$. Then we take
$\xi_{11}^{(0)}(0)=c_0,~\xi_{11}^{(0)'}(0)=0~\mbox{and}~
\xi_{01}^{(0)}(0)=c_1,~\xi_{01}^{(0)'}(0)=0$ for the exciton trial functions
of the discrete $(N=1)$ and continuous $(N=0)$ spectrum, respectively.
Comparing the corresponding functions of the discrete $\xi_{11}$ and continuous
$\xi_{01}$ states obtained by double integration of the set (\ref{E:set1})
using the chosen trial functions and the Coulomb functions determined
by eqs. (\ref{E:couldisc}) and (\ref{E:coulcont}), respectively, we arrive
at the set of equations

\begin{eqnarray}\label{E:setdouble}
\left.
\begin{array}{l}
c_0 Y_1 (\kappa) + c_1 \gamma_{01} =0;\\
c_1\left [\lambda_0 (\zeta)
-\frac{\pi}{1-{\rm e}^{-2\pi\zeta }}\cot \left(\Theta +\sigma  \right)   \right] + c_0\gamma_{01} =0;
\end{array}
\right \}
\end{eqnarray}
for the coefficients $c_0~\mbox{and}~c_1.$ In this equation
the functions $Y_1(\kappa)~\mbox{(\ref{E:Y}) and}~\lambda_0(\zeta)  $ (\ref{E:lambda}) define
the quantum numbers $\kappa$ and the phase $\Theta$ of the uncoupled discrete and continuous states,
respectively, while the parameter $\gamma_{01}$,
responsible for the inter-subband coupling, becomes

\begin{equation}\label{E:coupl}
\gamma_{01}=\frac{1}{\pi^2}\int_{-\frac{\pi}{2}}^{+\frac{\pi}{2}}d\varphi_e
\int_{-\frac{\pi}{2}}^{+\frac{\pi}{2}}d\varphi_h
\ln|\varphi_e - \varphi_h|\sin\varphi_h\sin\varphi_e = 0.386.
\end{equation}

The condition for solvability of equations (\ref{E:setdouble})

\begin{equation}\label{E:det}
Y_1 (\kappa)\left[\lambda_0 (\zeta)
-\frac{\pi}{1-{\rm e}^{-2\pi\zeta }}\cot \left(\Theta +\sigma  \right)   \right]-
\gamma_{01}^2=0
\end{equation}
establishes the relationship between the quantum numbers of the
discrete $(\kappa)$ and continuous $(\zeta)$ resonant states with the same
energy $E$

\begin{equation}\label{E:link}
E = \Delta_1\left(1  - \frac{q^2}{8\kappa^2}  \right) =
\Delta_0\left(1  + \frac{q^2}{8\zeta^2}  \right).
\end{equation}

In eq. (\ref{E:contmax}) for the coefficient of the exciton absorption
$\alpha^{(1)}(\omega)$ in the vicinity of the resonant energy $E$ (\ref{E:link})
we have to take $c_1 (s) = c_1 + c_0= c_1(1-\gamma_{01}Y_1 \kappa ^{-1} )$
calculated from the set (\ref{E:setdouble}), where $c_1$ is given by eqs. (\ref{E:zeros})
and (\ref{E:det}). It follows from equations (\ref{E:link}) and (\ref{E:lambda})
that $\zeta \simeq \frac{q}{\sqrt{3}}$ and
$\lambda_0 (\zeta) = \ln\frac{\pi}{\sqrt{3}} + C - \frac{1}{2} \simeq 0.68$.
Then, in the obtained equation (\ref{E:contmax}) for the coefficient
 $\alpha^{(1)}(\omega)$
we expand the function $Y_1(\kappa)$ (defined by (\ref{E:def}))
in the vicinity
of the energy $E_{1n}$ (\ref{E:discen}) obtained from the condition
(\ref{E:Y}). The absorption coefficient reads as follows

\begin{equation}\label{E:spectr}
\alpha_n^{(1)}(\omega) = \alpha_0\frac{4\pi p^2}{n_b \Delta_1 d}
\frac{f_n^{(1)}}{L}\Lambda_n (\hbar \omega -E_{1n} -\Delta E_{1n}),
\end{equation}
where

\begin{equation}\label{E:lor}
\Lambda_n (\hbar \omega -E_{1n} -\Delta E_{1n})=
\frac{\Gamma_{1n}}{2\pi\left[(\hbar \omega -E_{1n} -\Delta E_{1n})^2 + \frac{\Gamma_{1n}^2}{4} \right]}
\end{equation}
In eq. (\ref{E:spectr}) the specific oscillator strengths are the same
as those in eq. (\ref{E:discmax}). The following notation
for the resonant shift $\Delta E_{1n}$ and the resonant width $\Gamma_{1n}$
has been used

\begin{equation}\label{E:shift}
\Delta E_{1n} =
\frac{\lambda_0 \gamma_{01}^2 q^4 \Delta_1}{3\kappa_{1n}^3 \left(\frac{\partial Y}{\partial \kappa}\right)}~;
\end{equation}

\begin{equation}\label{E:width}
\Gamma_{1n} =
\frac{\gamma_{01}^2 q^3 \Delta_1}{\sqrt{3}\kappa_{1n}^3 \left(-\frac{\partial Y}{\partial \kappa}\right)}~;
\end{equation}
where

$$
\frac{\partial Y}{\partial \kappa}=
\left\{
\begin{array}{cl}
(-2\beta_{10}^2)^{-1}~;\quad n=0~;
\\
\\
(-\beta_{1n}^2)^{-1}~;\quad n=1,2,\ldots~;
\end{array}
\right.
$$
The quantum defects $\beta_{1n}=\kappa_{1n} - n$
can be calculated from eq. (\ref{E:def})

\section{Discussion}\label{S:discuss}

The exciton absorption spectrum
calculated in the single-subband approximation
consists of the sequence of the Rydberg $N$-series
of peaks of $\delta$-function type with intensities $|c_0(n)|^2$
(\ref{E:discmax}) and frequencies $\omega_n^{(N)}$ (\ref{E:discsp}) in the region
$\hbar\omega \leq \Delta_N$ and the branches of continuous absorption
$|c_1(s)|^2$ (\ref{E:contmax}) for $\hbar\omega \geq \Delta_N$.

\begin{figure}[htbp]
 \begin{center}

 \includegraphics[width= \linewidth]{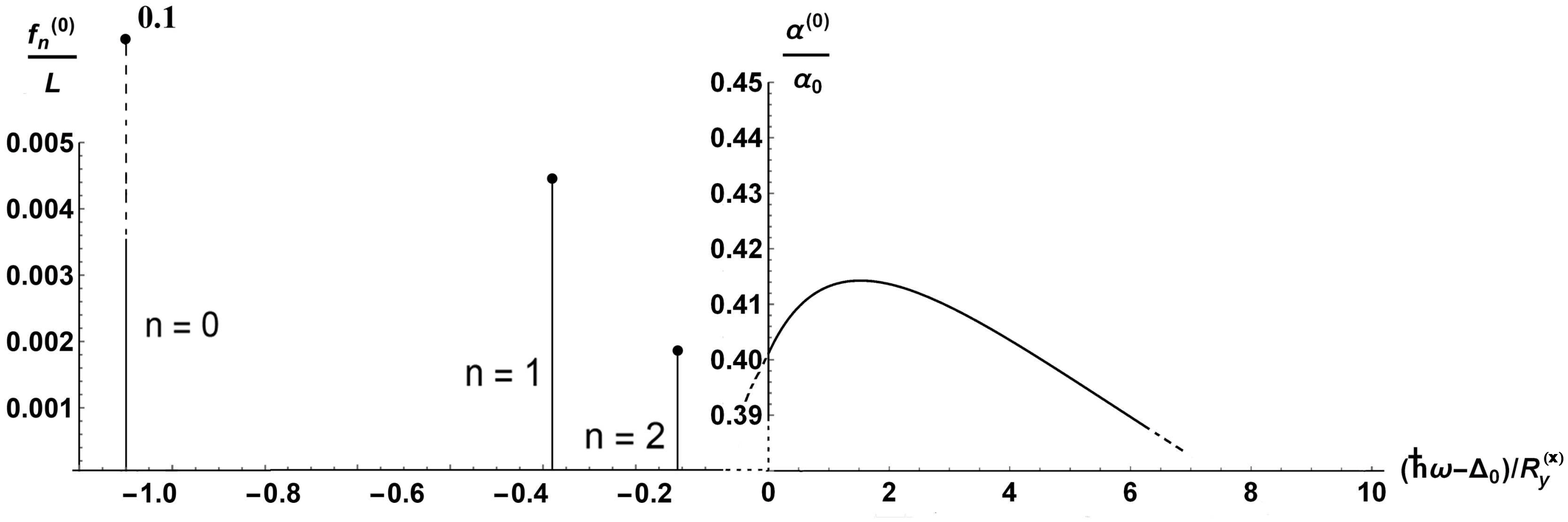}
 \end{center}
 \caption{\label{fig1} The specific oscillator strengths
$\frac{f_n^{(0)}}{L}$ (eq. (\ref{E:discmax})) and continuous absorption
$\frac{\alpha^{(0)}}{\alpha_0}$ (eq. (\ref{E:contmax})) associated with the ground
 $(N=0)$ energy gap $\Delta_0 = 0.69\,\mbox{eV}\,$ as a function of the frequency
 shift $\hbar \omega - \Delta_0 $ scaled with the exciton Rydberg constant
 $Ry^{\mbox{\tiny (x)}} = 13.8\,\mbox{meV}\,$ (\ref{E:discsp})
 in the AGNR of width $d=2\,\mbox{nm}\,$ placed on the sapphire substrate ($q \simeq 0.40$.) }
\end{figure}
All the Rydberg series $\alpha_n^{(N)}$, except for $\alpha_n^{(0)}$
adjacent to the ground size-quantized level $N=0$, overlap with the branches
of the continuous spectra, originating from the lower $N$-levels.
As a result only the ground series $N=0$ is formed by
transitions to the strictly discrete exciton states, while the others
$N\neq 0$ series are associated with transitions to the Fano resonant states,
induced by the inter-subband coupling between the overlapping
discrete $(n)$ and continuous $(s)$ exciton states, related to
various subbands \cite{Fano}. Thus in the multi-subband approximation
only the ground exciton Rydberg series $\alpha_n^{(0)}$ (\ref{E:absgen}) is
composed of $\delta$-function type peaks, while the others
$\alpha_n^{(N)}, N\neq 0$ consist of the absorption maxima of finite height
and nonzero Fano frequency width $\Gamma_{Nn}$, previously calculated
for the impurity electron in AGNR \cite{Monschm12} and for
low-dimensional semiconductor structures
(see \cite{Monschm09} and references therein). Below we focus
 on the exciton absorption spectrum
 associated with the ground size-quantized subband $N=0$
(see Fig.1).
The corresponding results coincide qualitatively with those calculated in the
single-subband approximation for the frequency regions
relevant to the subbands $N\neq 0$.

It follows from eqs. (\ref{E:absgen}) and (\ref{E:discmax}) that
the oscillator strengths of exciton peaks of the $\delta$-function type
at frequencies (\ref{E:discsp}) rapidly decrease as $\sim n^{-3}$
with increasing quantum number $n$. The intensities of the excited
peaks $n\geq 1$ scaled with the ground maximum $n=0$ obey the inequality
$\beta_{0n}^2\beta_{00}/ n^3 \ll 1$. Thus
as presented in Fig.1, optical absorption at $\hbar\omega < \Delta_0$ is
 practically concentrated in the region of transition to the ground exciton
 state $n=0$.
On narrowing the ribbon, the positions of peaks
(\ref{E:discsp}) $\omega_n \sim d^{-1}$ shift towards
higher frequencies, and their oscillator strengths
$f_n^{(0)}\sim |c_0(n)|^2$ increase in magnitude (Fig.2).

\begin{figure}[htbp]
 \begin{center}

 \includegraphics[width= 0.7\linewidth]{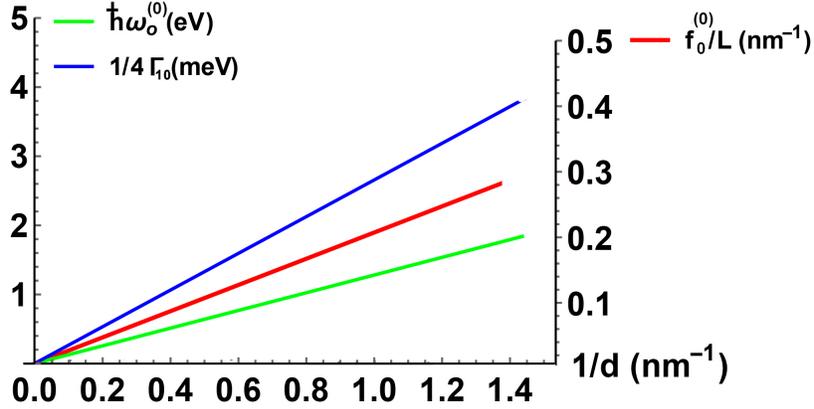}
  \end{center}
 \caption{\label{fig2} The ground exciton peak
position $\hbar \omega_0^{(0)}$ (eqs. (\ref{E:discsp}), (\ref{E:def})),
its specific oscillator strength $\frac{f_0^{(0)}}{L}$ (\ref{E:discmax}) and the
resonant width $\Gamma_{10}$ (eq. (\ref{E:width})) versus the inverse ribbon
width $d^{-1}$ of the AGNR situated on a sapphire substrate ($q \simeq 0.40$.) }
\end{figure}
With decreasing ribbon width $d$ the distance
$| \hbar\omega_n^{(0)} - \Delta_0 |\sim d^{-1} $ between
the frequency positions of peaks $\omega_n^{(0)}$
and the edge of continuous absorption $\Delta_0 / \hbar$ increases,
which in turn makes the narrow AGNR preferable candidates
for the experimental study of a discrete exciton spectrum.
The dependencies of the binding energy of the ground exciton
state $E_{00}^{(b)} = \Delta_0 - E_{00}$ calculated
from eq. (\ref{E:discen}), of
the corresponding specific oscillator strength
$f_0^{(0)}/L$ and of the width $\Gamma_{10}$
on the inverse ribbon width $d^{-1}$
and on the exciton interaction strength $q$ are shown in Fig.3.

\begin{figure}[htbp]
 \begin{center}

 \includegraphics[width= 0.9\linewidth]{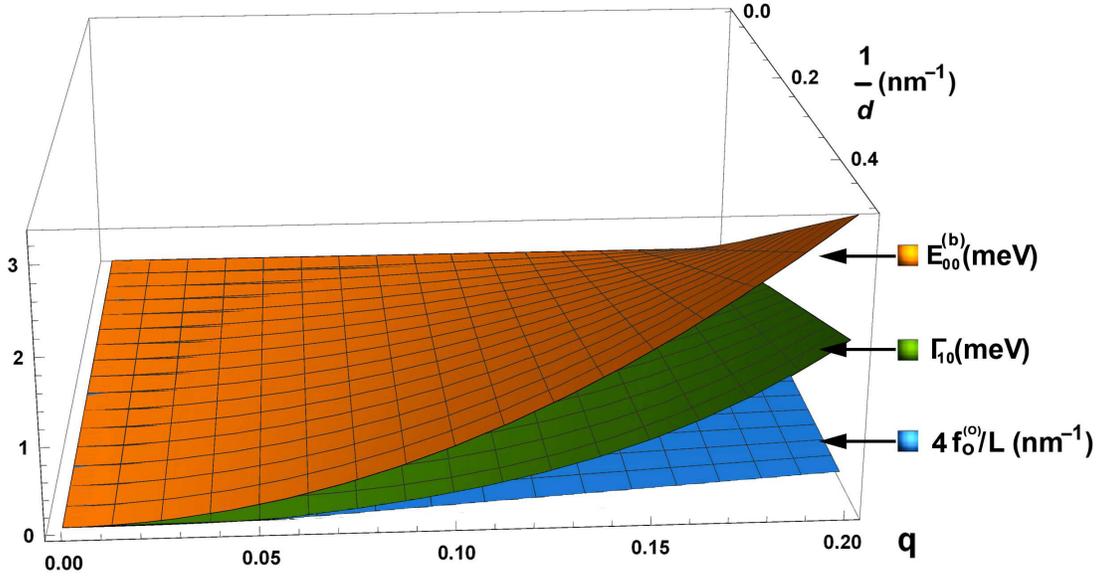}
  \end{center}
 \caption{\label{fig3}  The dependencies
of the binding energy $E_{00}^{(b)}$ (eq.(\ref{E:discen}))
of the ground discrete exciton state, of
the corresponding
specific oscillator strength $\frac{f_0^{(0)}}{L}$ (eq. (\ref{E:discmax}))
and of the resonant width $\Gamma_{10}$ (eq. (\ref{E:width}))
of the ground quasi-discrete exciton state
on the inverse ribbon width $d^{-1}$ and on the dimensionless
exciton parameter $q$ defined in eq. (\ref{E:adiab}). }
\end{figure}
In the continuous spectral region $\hbar\omega\geq  \Delta_0$
(Fig.1) the exciton absorption $\alpha^{(0)}(\omega)$ (\ref{E:contmax})
is the fundamental absorption
 $\alpha^{(0)}(\omega) \sim \zeta \sim (\hbar\omega  - \Delta_0)^{-\frac{1}{2}} $
modified by the 1D Zommerfeld factor $Z(\zeta)$,
taking into account the exciton attraction of the electron and hole.
For the frequencies positioned far away from the edge $(\zeta \ll 1)$,
the exciton interaction has little effect $(Z\simeq 1)$
on the fundamental absorption. In the vicinity of the edge
$(\zeta\rightarrow \infty~,\,Z\rightarrow (2\pi\zeta)^{-1})$
the exciton factor modifies strongly the fundamental absorption
compensating the singularity $\sim\zeta$ and providing the finite
absorption $\alpha^{(0)}(\omega) $ (\ref{E:edge1}) at the edge
and the linear addition $\Delta\alpha^{(0)}\sim \zeta^{-2}$
nearby the edge. The main characteristics of the optical spectrum,
represented in Fig.1, namely the radical redistribution of the absorbed
radiation in favor of the ground peak and the disappearance of the
singularity of the 1D density of states, are the
common signatures of the 1D exciton absorption. Earlier
this was found for 1D excitons in bulk semiconductors subject
to a strong magnetic field \cite{Hashow} and in semiconductor
quantum wires \cite{Monschm09}. Recently an analogous result was
obtained by Portnoi \emph{et al.} \cite{Port} employing a
1D quasi-Coulomb potential for the calculation of the exciton absorption
spectrum in narrow-gap carbon nanotubes.

It follows from eqs. (\ref{E:absgen}) and (\ref{E:spectr})
derived in the double-subband approximation that the inter-subband
coupling modifies the quasi-Rydberg exciton series adjacent
to the excited size-quantized energy levels $N\neq 0$, thereby
replacing the $\delta$-function type optical peaks for which
$\alpha_n^{(N)}(\omega) \sim f_n^{(N)}
\delta (\hbar \omega - E_{Nn} )$, by
those of the Lorentzian form with
finite maxima
$\Lambda_{\mbox{max}} \sim \frac{2}{\pi \Gamma_{Nn}}$
and nonzero width $\Gamma_{Nn}$. The resonant peaks are red-shifted
by an amount $\Delta E_{Nn}$.
Since the resonant shifts $\Delta E_{Nn}$ and widths $\Gamma_{Nn}$
can be qualitatively described by eqs. (\ref{E:shift})
and (\ref{E:width}), respectively, we take

$$
\Gamma_{Nn} \sim E_{Nn}^{(b)}\frac{q}{\kappa_{Nn}};~
\Delta E_{Nn}\sim - E_{Nn}^{(b)}\frac{q^2}{\kappa_{Nn}};
~E_{Nn}^{(b)}\sim \Delta_N \frac{q^2}{\kappa_{Nn}^2};\quad
\kappa_{Nn} =n + \beta_{Nn}.
$$

Note firstly that the resonant shifts $\Delta E_{Nn}\sim \Delta_N q^4$
are much less than the resonant widths $\Gamma_{Nn} \sim \Delta_N q^3$
at $q\ll 1,~ (\Delta E_{Nn} \ll \Gamma_{Nn})$ and secondly the resonant
shifts do not change the discrete character of the exciton energy
spectrum at $E \leq \Delta_N$, while the resonant widths
are the characteristics of the continuous spectrum accounting for the
finite life-times $\tau_{Nn} = \hbar / \Gamma_{Nn}$ of the
exciton states. We point out that the relationship
between $\Delta E_{Nn}$ and $\Gamma_{Nn}$ strongly depends
on the dimension of the structure. For 1D structures, namely
AGNR, bulk semiconductors subject to strong magnetic fields
\cite{Zhilmak} and quantum wires \cite{Monschm09},
$\Delta E_{Nn} \ll \Gamma_{Nn}$, while for 2D systems
such as QW \cite{Monschm05} and superlattices \cite{Monschm07},
$\Delta E_{Nn} \gg \Gamma_{Nn}$. Eq. (\ref{E:width}) and Figs. 2 and 3
show that the resonant widths $\Gamma \sim E^{(b)} q,
~q\simeq \frac{d}{a_{\mbox{\tiny x}}(d)} $
increase with decreasing ribbon width $d$. This dependence
is opposite to that in a QW, in which the narrowing of the well decreases
the resonant widths
$ \Gamma^{(w)}\sim E_{w}^{(b)}q_w^4;~q_w =\frac{d}{a_{\mbox{\tiny x}}^{(w)}} $
\cite{Monschm05}. This is because the dependencies of the binding energies
$E^{(b)}$ and the adiabaticity parameter $q$ are completely different
for the AGNR and QW. In the AGNR $E^{(b)} \sim \Delta_N \sim d^{-1}$
increases on narrowing the ribbon, while
$q\sim \frac{d}{a_{\mbox{\tiny x}}(d)}$ does not depend on the ribbon width $d$.
In a QW the binding energy
$E_w^{(b)}\sim Ry_w^{\mbox{\tiny (x)}}\sim \frac{1}{a_{\mbox{\tiny x}}^{(w)2}}$
and the Bohr radius $a_{\mbox{\tiny x}}^{(w)}$ are independent of the well width $d$,
while with decreasing $d$ the parameter
$q_w \sim \frac{d}{a_{\mbox{\tiny x}}^{(w)}}\sim d$ decreases as well.
The analogous conclusion
is valid for the resonant widths of the exciton states
in a semiconductor superlattice \cite{Monschm07}. Exciton peaks associated
with the transitions to the excited resonant states $n=1,2,\ldots$
are much narrower than that corresponding to the ground state $n=0$ with
$\Gamma_{1n}/\Gamma_{10}\simeq \beta_{1n}^2 \beta_{10}/2n^3 \ll 1. $
This strongly affects the relationship between the maximum values
of peaks of absorption (\ref{E:spectr}) at
$\hbar \omega = E_{1n} + \Delta E_{1n}$

$$
\alpha_{n\mbox{\tiny max}}^{(1)}=\alpha^{(0)}
\frac{8p^2 f_n^{(1)}}{n_b \Delta_1 d L \Gamma_{1N}}
$$
related to the ground $n=0$ and excited $n=1,2.\ldots$ exciton
states. It follows from Eqs. (\ref{E:discmax}) and (\ref{E:width})
for the oscillator
strengths and the resonant widths, respectively, that
$\frac{\alpha_{n\mbox{\tiny max}}^{(1)}}{\alpha_{0\mbox{\tiny max}}^{(1)}}=1.$
Thus in contrast to the exciton series adjacent to the ground
subband $N=0$ for which the ground exciton peak $n=0$ significantly
exceeds in intensity the excited ones $n=1,2,\ldots$
$\left(\frac{f_n^{(0)}}{f_0^{(0)}}=
 \frac{\beta_{0n}^2 \beta_{00}}{2n^3}\ll 1\right)$ the series related
 to the excited subbands $N\neq 0$ consist of exciton peaks
 with comparable intensities.

In view of possible future experiments, we estimate
the edges of the inter-subband optical absorption
$\hbar\omega^{(N)}= \Delta_N$, the exciton binding energies
$E_{Nn}^{(\mbox{b})}=Ry^{\mbox{\tiny(x)}}/(n+\beta_{Nn})^2$
and the specific oscillator strengths $f_n^{(N)}/L$
of the transitions to the exciton states for the AGNR
placed on a sapphire substrate $(\varepsilon \simeq 10, q \simeq 0.40)$.
The latter
is preferable compared to a $\mbox{SiO}_2$ substrate
$(\varepsilon \simeq 3.9)$,
which provides a smaller screening of the
electron-hole attraction and larger value of the parameter $q$.
The energy
gaps, the binding energies and the oscillator strengths
determine the exciton peak positions (\ref{E:discsp}) and their
intensities (\ref{E:discmax}), respectively. The edges of
optical absorption calculated from eq. (\ref{E:tren}) $\Delta_N =2\varepsilon_N$
for $N=0,1,-1,2,-2,3~;\tilde{\sigma}=\frac{1}{3}$ for the
ribbon consisting of 55 dimers $(d\simeq 6.6~\mbox{nm})$ and those
presented by Sasaki \emph{et al.} (Fig.3 in Ref. \cite{Sas} )
are given in Table 1. For the ribbon of width $d \simeq 1.7~\mbox{nm}$
(15 dimers) eq. (\ref{E:tren}) and Sasaki \emph{et al.} calculations lead to
amounts
$\Delta_0\simeq 0.80~\mbox{eV~\,and}\quad \Delta_0^{*}\simeq 0.66~\mbox{eV}$,
respectively.

\begin{table}[h]

\begin{center}
\begin{tabular}{||l|l|l|l|l|l|l||}
\hline
 $N$        &$0$&$1$&$-1$&$2$&$-2$&$3$ \\ \hline
 \hline
$\Delta_N$&     0.20     & 0.40 & 0.80 & 1.0 & 1.4 & 1.6  \\ \hline
$\Delta_N^{*}$& 0.20     & 0.40 & 0.80 & 0.9 & 1.4 & 1.5  \\ \hline
\end{tabular}
\end{center}
 \caption{The edges of the
optical absorption calculated from eq. (\ref{E:tren}) $\Delta_N =2\varepsilon_N$
for $N=0,1,-1,2,-2,3~;\tilde{\sigma}=\frac{1}{3}$ and those $\Delta_N^{*}$
given in Ref. \cite{Sas}.}
\end{table}


Son \emph{et al.} \cite{Son} calculated the energy gap
$\Delta_0^{*}=0.50~\mbox{eV}$ for the ribbon of width $2~\mbox{nm}$
with hydrogen passivated edges. On excluding the $12\% $ gap reduction
caused by the passivation resulting in $\Delta_0^{*}=0.57~\mbox{eV}$
we compare the later with our optical edge found from eq. (\ref{E:tren})
 $\Delta_0=0.69~\mbox{eV}$. As expected, our data for $\Delta_N$ and
 the values of $\Delta_0^{*}$ presented in Refs. \cite{Sas} and \cite{Son} are
 in very good agreement (see Table 1) for a relatively wide ribbon
 $(d \geq  6~\mbox{nm})$, but not for narrow samples
 $ (d \leq  2~\mbox{nm})$. The reason therefore is that
 the analytical Dirac equation
 method treating the ribbon as a continuous medium is good for wide
 ribbons, while the numerical tight-binding approximation,
 which takes into account the discrete atomic structure of a ribbon, provides
 more accurate results for narrow samples. The calculation of the
 binding energy
$E_{00}^{(\mbox{b})}$ of the ground exciton state $(n=0)$
located within the ground energy gap $(N=0)$ using
eqs. (\ref{E:discsp}), (\ref{E:discen}) and (\ref{E:def}) for an AGNR
of width $d=1~\mbox{nm}$ placed on a sapphire substrate yields a value
$E_{00}^{(\mbox{b})} \simeq 30~\mbox{meV}$.
The value of the specific oscillator strength
$\frac{f_0^{(0)}}{L}\simeq \frac{q}{2 \beta_{00} d}$
calculated using eqs. (\ref{E:discmax})
and (\ref{E:def}) is equal to
$\frac{f_0^{(0)}}{L}\simeq 0.20~\mbox{nm}^{-1} $. The relatively small
values of the binding energy and specific oscillator strength are the consequences
of high dielectric constant $\varepsilon$ of the substrate,
which ensures the condition $q << 1$ for the
adiabatic approximation. Any detailed
quantitative comparison of our results with those obtained
earlier by both analytical and numerical methods is problematic
because the latter were obtained under different conditions, e.g.,
for a suspended AGNR $(q=2.2)$ \cite{Zhu, Jia}
and hydrogen-passivated edges \cite{Prez} or with an unspecified dielectric
constant $\epsilon_{\mbox{\tiny eff}}(\vec{r})$ \cite{Mohamm}.

We estimate the binding energy
$E_{10}^{(\mbox{b})} = \Delta_1 - \hbar \omega_0^{(1)}$
of the ground Fano-resonant exciton state $n=0,N=1$,
using eq. (\ref{E:discsp})
for an AGNR
of width $d= 2~\mbox{nm}$ placed on the sapphire substrate.
The quantum defect $\beta_{10}$
determining the binding energy $E_{10}^{(\mbox{b})}$
has been calculated from eqs. (\ref{E:Y}) and (\ref{E:def}).
For the strictly discrete ground series $N=0$ of the considered AGNR,
$E_{00}^{(\mbox{b})}=15~\mbox{meV}$, which exceeds the
binding energy $E_{10}^{(\mbox{b})}\simeq 3.8~\mbox{meV}$.
The latter is in line with the
conclusion made in Ref. \cite{Zhilmon} that the exciton series
relevant to the $N$-subband become markedly suppressed
with increasing subband index $N$.

The width $\Gamma_{10}$ (\ref{E:width}) and the maximum absorption coefficient
$\alpha_{0\mbox{\tiny max}}^{(1)}$ (\ref{E:spectr}) with $n_b \simeq 1.8$
possess the values $\Gamma_{10} \simeq 5.75~\mbox{meV}
~\mbox{and}~\alpha_{0\mbox{\tiny max}}^{(1)}= 0.214$.
Clearly, the exciton Fano resonances with the lifetimes
$\tau =\hbar/\Gamma_{10}=0.12~\mbox{ps}$
can be detected in optical
experiments with narrow AGNR. The considerable enhancement of
exciton optical absorption
in quasi-1D AGNR with respect to
excitonless 2D graphene layers, for which $\alpha_0 = 2.3\cdot10^{-2}$
can be used in optoelectronics and applied optics. We believe
that the proposed analytical approach will be useful for both
theoretical studies and practical
applications of the scalable exciton effects in the AGNR.

\section{Conclusions}\label{S:concl}

We have developed an analytical approach  to the problem of the
exciton absorption in the narrow gap armchair graphene
nanoribbon. The ribbon width is taken to be much less
than the exciton Bohr radius. This adiabatic criterion allows
us to solve analytically the two-body 2D Dirac equation,
 describing the interacting massless electron-hole pair
and then to calculate the optical
absorption coefficient in an explicit form.
With the coupling between the different subbands taken into account,
the Fano resonances appear instead of strictly discrete exciton states.
The exciton spectrum is a sequence
of Rydberg series of strictly discrete
or broadened resonant peaks positioned within the gaps
determined by the electron-hole size-quantized energy levels
and continuous bands, branching from the tops of the gaps.
The intensities, frequency widths, and blue shifts of exciton peaks increase
on narrowing the ribbon.
At the edges, the exciton effect
eliminates the singularities in the fundamental absorption.
Our analytical results are
in good agreement with those obtained by using other
theoretical approaches, in particular, with the results of numerical studies.
The expected experimental values are estimated for concrete
AGNR.

\section{Acknowledgments}\label{S:Ackn}
The authors are grateful to D.~Turchinovich and V.~Bulychev
for useful discussions and to T.~Fedorova for technical assistance.

\end{document}